\documentclass[a4paper]{jpconf}
\usepackage{graphicx}
\begin{document}
\title{Collective enhancement of nuclear state densities by the shell model Monte Carlo approach}

\author{C~\"{O}zen$^{1}$, Y~Alhassid$^{2}$ and H~Nakada$^{3}$}

\address{$^1$ Faculty of Engineering and Natural Sciences, Kadir Has University, Istanbul 34083, Turkey \\
$^2$ Center for Theoretical Physics, Sloane Physics Laboratory, Yale University, New Haven, CT 06520, USA \\
$^3$ Department of Physics, Graduate School of Science,Chiba University, Inage, Chiba 263-8522, Japan}

\ead{cem.ozen@khas.edu.tr}

\begin{abstract}
The shell model Monte Carlo (SMMC) approach allows for the microscopic calculation of statistical and collective properties of heavy nuclei using the framework of the configuration-interaction shell model in very large model spaces.  We present recent applications of the SMMC method to the calculation of state densities and their collective enhancement factors in rare-earth nuclei.  
\end{abstract}

\section{Introduction}
Collective states constitute a significant part of the spectra of heavy nuclei up to moderate excitation energies and they are often described by bands that are built on top of intrinsic states. 
However, the contribution of these collective states to the nuclear state density is difficult to calculate microscopically, and is often described by phenomenological collective enhancement factors~\cite{RIPL}. The challenge of computing the collective enhancement factors microscopically can be overcome using the shell model Monte Carlo (SMMC) method~\cite{Lang1993,Alhassid1994,Koonin1997,Alhassid2001}. This method is particularly suitable for the calculation of thermal and statistical properties of nuclei.

Here we discuss recent applications of the SMMC method to the isotopic families of even-even samarium and neodymium nuclei~\cite{Ozen2013a,Alhassid2013a}, which exhibit a crossover from vibrational to rotational collectivity. In particular, we present results for the state densities and their collective enhancement factors.

\section{The shell model Monte Carlo (SMMC) approach}
\label{sec-smmc}
The SMMC method is based on the Hubbard-Stratonovich (HS) transformation~\cite{HS-trans} to express the Gibbs operator $e^{-\beta H}$ of a nucleus (described by a Hamiltonian $H$ at inverse temperature $\beta=1/T$) as a superposition
of one-body propagators of non-interacting nucleons moving in external auxiliary fields $\sigma(\tau)$ that depend on imaginary time $\tau$ ($0\leq \tau \leq \beta$)
\begin{eqnarray}
 e^{-\beta H} = \int D[\sigma] G_\sigma U_\sigma \;.
\end{eqnarray}
Here $G_\sigma$ is a Gaussian factor and $U_\sigma$ describes a one-body propagator associated with a given set of auxiliary fields $\sigma$. Subsequently, the thermal expectation value of an observable $O$ at inverse temperature $\beta$ can be written in the HS representation as
\begin{eqnarray}
\label{observable}
\langle O\rangle = {\Tr \,( O e^{-\beta H})\over  \Tr\, (e^{-\beta H})} = {\int D[\sigma] W_\sigma \Phi_\sigma \langle O \rangle_\sigma
\over \int D[\sigma] W_\sigma \Phi_\sigma} \;,
\end{eqnarray}
where  $\langle O \rangle_\sigma = \Tr \,(O U_\sigma)/ \Tr\,U_\sigma$ is the thermal expectation value of the observable in a given configuration of the auxiliary fields $\sigma$. Since the numbers of neutrons and protons are fixed for a given nucleus, all traces in Eq.~(\ref{observable})
are evaluated in the canonical ensemble. Defining a positive-definite function $W_\sigma = G_\sigma |\Tr\, U_\sigma|$ and the associated Monte Carlo sign
$\Phi_\sigma = \Tr\, U_\sigma/|\Tr\, U_\sigma|$, auxiliary-field configurations $\sigma_k$ are sampled according to $W_\sigma$, and the expectation value in (\ref{observable}) is then estimated from
 $\langle  O\rangle \approx  {\sum_k
  \langle O \rangle_{\sigma_k} \Phi_{\sigma_k} / \sum_k \Phi_{\sigma_k}}$.
  
\section{Collective enhancement in the state densities of heavy nuclei}
\label{sec-coll}
Nuclear collectivity is a ubiquitous phenomenon in heavy nuclei. Various types of collectivity,  
such as vibrational collectivity and rotational collectivity, are observed and are well described by phenomenological  models. However,  a microscopic description of nuclear collectivity using the configuration-interaction (CI) shell model is a challenging task. In particular, CI shell model calculations of heavy nuclei require model spaces that are many orders of magnitude larger than model spaces that can be treated by conventional diagonalization methods.
The SMMC was successfully applied to describe the rotational character of the rare-earth nucleus $^{162}$Dy~\cite{Alhassid2008} in the framework of a truncated spherical shell model space. Recently, using the same approach and model space, the crossover from vibrational to rotational collectivity in families of even-even samarium and neodymium isotopes was also reproduced~\cite{Ozen2013a,Alhassid2013a}. Here we discuss the state densities of these nuclei and the associated collective enhancement factors.    
 
The single-particle model space consists of the $50-82$ major shell orbitals plus the $1f_{7/2}$ orbital for protons, and the $80-126$ major shell orbitals plus the $0h_{11/2}$ and $1g_{9/2}$ orbitals for neutrons. The bare single-particle energies of the orbitals were chosen so they reproduce the Woods-Saxon energies in the spherical Hartree-Fock approximation. The effective two-body interaction consists of monopole pairing interaction terms for protons and neutrons, and multipole-multipole interaction with quadrupole, octupole and hexadecupole terms. The calculations were performed using an SMMC code in the proton-neutron formalism~\cite{Alhassid2008}.
 
\subsection{State Densities}
\label{sec-dens}
The SMMC method has proven to be a powerful technique for the calculation of state densities~\cite{NA1997,Alhassid1999}. In the SMMC approach, the thermal energy is calculated as the expectation value of the many-body Hamiltonian at an inverse temperature $\beta$. The canonical partition function is computed by integrating the thermal energy with respect to $\beta$, and the state density is obtained by inverting the Laplace transform of the canonical partition function in the saddle-point approximation.

\begin{figure} [h!]
\centering
\includegraphics[width=0.9\columnwidth,clip]{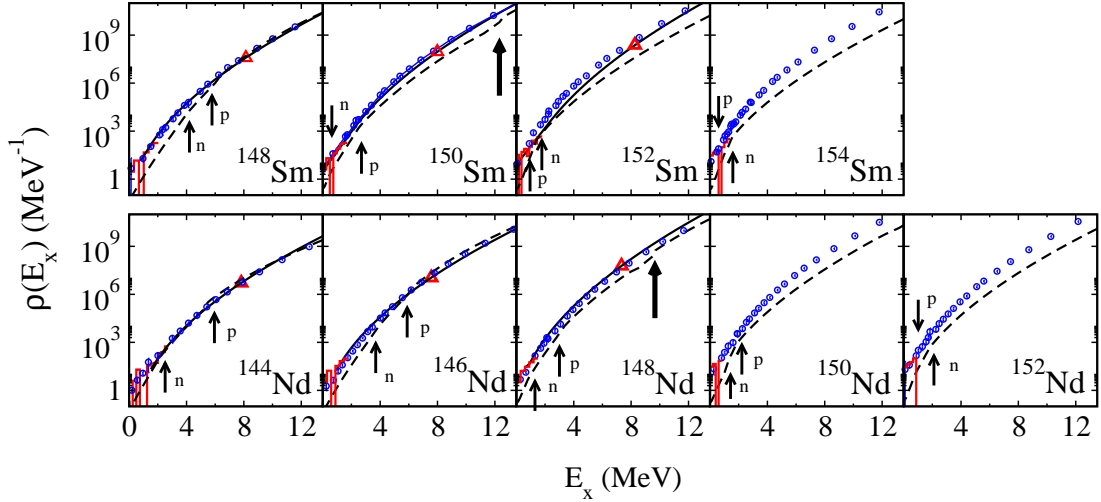}
\caption{State densities in the even-even $^{148-154}$Sm and $^{144-152}$Nd isotopes.
The SMMC densities (circles) are compared with level counting data (histograms) and
neutron resonance data (triangles) when available. Also shown are the BBF state densities (solid lines), which are parametrized using the experimental data, and the HFB densities (dashed lines). The thin arrows indicate the neutron and proton pairing phase transitions, and the thick arrows indicate the shape phase transitions. Adapted from Refs.~\cite{Ozen2013a,Alhassid2013a}.}
\label{fig-rho-even}
\end{figure}

In Fig.~\ref{fig-rho-even} we show the SMMC state densities calculated for the even-even $^{148-154}$Sm and $^{144-152}$Nd isotopes. We compare the SMMC results (circles) with the experimental state densities at low energies using level counting data (histograms),  and at the neutron resonance energy (when available) assuming the spin cutoff model with rigid-body moment of inertia (triangles). We also show the BBF state densities (solid lines) whose parameters (i.e., the single-particle level density parameter and the backshift parameter) are determined from level counting and neutron resonance data. The SMMC results are also compared with the densities calculated by the finite-temperature Hartree-Fock-Bogoliubov (HFB) approximation (dashed lines). The observed enhancement of the SMMC densities over the HFB densities can be attributed to the presence of collectivity in these nuclei, since  collective excitations are included in the SMMC approach but are absent in the HFB densities. The HFB densities also display ``kinks'' at energies that correspond to the phase transitions associated with the breaking of the proton and neutron pairs (thin arrows) and with the transitions from deformed to spherical shapes (thick arrows). These shape transitions occur in all nuclei except  $^{148}$Sm, $^{144}$Nd and $^{146}$Nd, which are spherical in their ground states.  The shape transitions in $^{152}$Sm, $^{154}$Sm, $^{150}$Nd and $^{152}$Nd occur at excitation energies that are higher than those displayed in Fig.~\ref{fig-rho-even}. They can be seen, however,  in Fig~\ref{fig-collective}.

\subsection{Collective Enhancement}
\label{sec-enh}

The microscopic calculation of collective effects in the state density of heavy nuclei has been a major challenge. Such effects are usually described in terms of phenomenological vibrational and rotational collective enhancement factors~\cite{RIPL}. 

\begin{figure}
\centering
\includegraphics[width=0.9\columnwidth,clip]{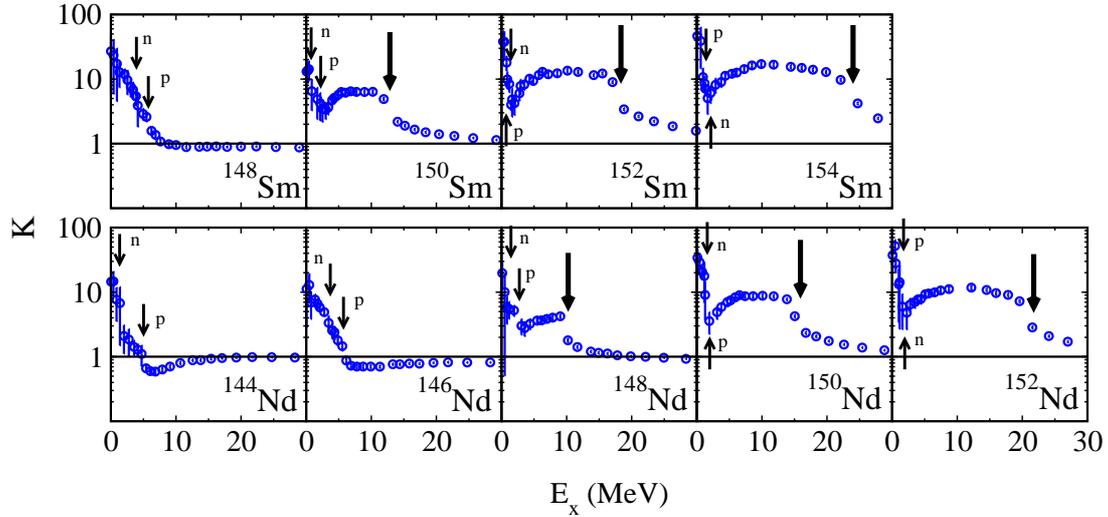}
\caption{Collective enhancement factor K versus excitation energy $E_x$ in the even-even $^{148-154}$Sm and $^{144-152}$Nd isotopes. The pairing and shape transition energies are shown by the thin and thick arrows, respectively. Adapted from Refs.~\cite{Ozen2013a,Alhassid2013a}.}
\label{fig-collective}
\end{figure}

Recently we introduced~\cite{Ozen2013a,Alhassid2013a} the ratio of the SMMC and the HFB densities, i.e.,  $K=\rho_\mathrm{SMMC}/\rho_\mathrm{HFB}$, as a microscopic measure of the collective enhancement factor.  In Fig.~\ref{fig-collective}, we show $K$ as a function of the excitation energy $E_x$ for
the above families of samarium and neodymium isotopes. In the spherical nuclei $^{148}$Sm, $^{144}$Nd, and $^{146}$Nd, collectivity is predominantly vibrational and is observed to be lost completely (i.e., $K \sim 1$) above the pairing transition energies. The other samarium and neodymium isotopes are deformed, and collectivity persists at higher excitation energies than the pairing transition energies where the enhancement is due to rotational collectivity.  Above the shape transition energy, the collective enhancement factor decays to $K\sim 1$ since spherical nuclei can no longer support rotational bands. 

\section{Conclusion}
\label{sec-concl}

We discussed recent SMMC applications to the even-even $^{148-154}$Sm and $^{144-152}$Nd isotopes. 
The total SMMC state densities were shown to be in very good agreement with experimental data. We also extracted a 
microscopic measure of the collective enhancement factor defined by the ratio of the SMMC and HFB densities.
We observed  that the decay of vibrational and rotational collectivity with excitation energy correlates with the pairing and shape phase transitions, respectively. 

\ack
This work was supported in part by the U.S. Department of Energy Grant No.~DE-FG02-91ER40608, and by the Grant-in-Aid for Scientific Research (C) No.~25400245 by the JSPS, Japan.  The research presented here used resources of the National Energy Research Scientific Computing Center, which is supported by the Office of Science of the U.S. Department of Energy under Contract No.~DE-AC02-05CH11231.  It also used resources provided by the facilities of the Yale University Faculty of Arts and Sciences High Performance Computing Center.

\end{document}